\documentclass[conference]{IEEEtran}
\IEEEoverridecommandlockouts

\usepackage{braket}
\usepackage{cite}
\usepackage{amsmath,amssymb,amsfonts}
\usepackage{algorithmic}
\usepackage{graphicx}
\usepackage{textcomp}
\usepackage{xcolor}
\def\BibTeX{{\rm B\kern-.05em{\sc i\kern-.025em b}\kern-.08em
    T\kern-.1667em\lower.7ex\hbox{E}\kern-.125emX}}
\begin{document}

\title{Analysis of a 3D Integrated Superconducting Quantum Chip Structure}

\author{\IEEEauthorblockN{James Saslow}
\IEEEauthorblockA{\textit{Department of Electrical Engineering} \\
\textit{San José State University}\\
San José, USA \\
james.saslow@sjsu.edu}
\and
\IEEEauthorblockN{Hiu Yung Wong}
\IEEEauthorblockA{\textit{Department of Electrical Engineering} \\
\textit{San José State University}\\
San José, USA \\
hiuyung.wong@sjsu.edu}
}

\maketitle

\begin{abstract}
This work presents a combined analytical and simulation-based study of a 3D-integrated quantum chip architecture. We model a flip-chip-inspired structure by stacking two superconducting qubits fabricated on separate high-resistivity silicon substrates through a dielectric interlayer. Utilizing \emph{rigorous} Ansys High-Frequency Structure Simulator (HFSS) simulations and analytical models from microwave engineering, we evaluate key quantum metrics such as eigenfrequencies, Q-factors, decoherence times, anharmonicity, cross-Kerr, participation ratios, and qubit coupling energy to describe the performance of the quantum device as a function of integration parameters. The integration parameters include the thickness and the quality of the dielectric interlayer. For detuned qubits, these metrics remain mostly invariant with respect to the substrate separation. However, introducing dielectric interlayer loss decreases the qubit quality factor, which consequentially degrades the relaxation time of the qubit. It is found that for the structure studied in this work, the stacked chip distance can be as small as $0.5 \text{mm}$. These findings support the viability of 3D quantum integration as a scalable alternative to planar architectures, while identifying key limitations in qubit coherence preservation due to lossy interlayer materials.
\end{abstract}


\section{Introduction}

\subsection{Motivation}

Three-dimensional (3D) integration of quantum chips is an increasingly active field of research, offering solutions to the spatial limitations present in two-dimensional (2D) planar chip architectures. A common 3D implementation strategy involves maintaining control electronics on one planar chip and superconducting qubits on another, then bonding the two chips together with an indium bump bonding flip-chip technique \cite{b13}. 
The control electronics chip routes signals through the bulk of the substrate using through-silicon vias (TSV's), reducing wire congestion on the surface of the chip \cite{b14}.

DWave, by contrast, manifests 3D integration via a stacked planar chip approach, distributing superconducting qubits across multiple chips \cite{b13}. DWave utilizes a Chimera topology, composed of $K_{4,4}$
unit cells that contain four horizontal qubits coupled to four vertical qubits \cite{b12}. This paper has a similar integration scheme.

Spatial constraints are a common struggle of typical 2D quantum chip layouts, usually resulting in limited connectivity between qubits and over-cluttering of electronics. 3D integration motivates the possibility of coupling qubits in the $x$, $y$, and $z$ directions via stacked planar geometries. However, such design choices introduce their own set of challenges to mitigate, such as determining an optimal plate separation to ensure sufficient coupling while minimizing cross-talk, as well as the presence of unwanted effects in the dielectric such as the loss tangent. This paper undergoes a rigorous simulation study to carefully model the advantages and consequences of implementing a 3D quantum integration design with lossy materials.

\subsection{Paper Contents and Organization}


First, in Section~\ref{sec: II. Design and Device Specifications}, we discuss the design specifications of our quantum device and the software tools we use for 3D design. Then, in Section~\ref{sec: III Analytical Calculations and HFSS Simulations of the 3D-Integrated Chip}, we perform a parametric Driven Modal simulation method in HFSS for impedance line matching to determine an accurate coplanar waveguide (CPW) characteristic impedance. We also do an eigenmode simulation to extract the entire device's resonant modes and quality factors. The optimized mesh from the eigenmode simulation is imported into our Driven Modal setup for a subsequent scattering analysis, ensuring accurate, physical, and consistent results. 
In Section~\ref{sec: IV Effect of dielectric interlayer thickness}, we investigate how substrate separation influences the qubit relaxation time, $T_{1}$, by performing a parametric eigenmode analysis in HFSS. Finally, in Section~\ref{sec: V Varying the loss tangent of the dielectric}, we insert a loss tangent in a low-$\epsilon_{r}$ dielectric (assumed to be $1$) positioned in between the two silicon substrates. Our HFSS eigenmode simulations reveal that the presence of dielectric loss tangent substantially degrades $T_{1}$ times and introduces an additional dephasing induced by the lossy material.

\section{Design and Device Specifications}
\label{sec: II. Design and Device Specifications}

\subsection{Tools and Software}

We utilize Qiskit Metal (version 0.1.5) for superconducting quantum circuit layout \cite{b10}, HFSS for electromagnetic simulation, and pyEPR (version 0.9.0) \cite{b6} for extracting quantum metrics of our circuits. Qiskit Metal and pyEPR run in a Python 3.9.0 environment. HFSS simulation is conducted under Ansys Electronics Desktop 2024 R2. Qiskit Metal generates a layout and sends it to Ansys to create a 3D rendering. In Ansys HFSS, we can perform scattering and eigenmode simulations to understand the behavior of our circuit within a classical electrodynamics context. We utilize the pyEPR package to extrapolate quantum-based metrics such as anharmonicity, cross-Kerr, and junction participation based on the field distributions in HFSS.

\subsection{3D Integrated Structure Design and Construction}

\begin{figure}[tb]
\centerline{\includegraphics[width=0.5\textwidth]{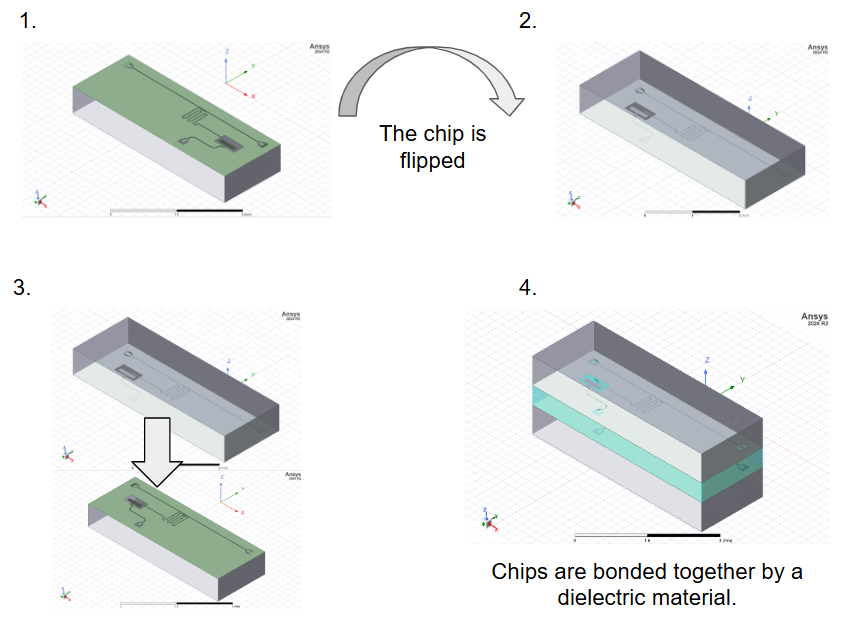}}
\caption{The 3D integration flip-chip scheme studied in this paper. 1) We start with our top high-resistive silicon substrate with a laid-out quantum circuit assumed to be fabricated with niobium and an Al/AlOx/Al Josephson junction. 2) Then, we flip the top substrate such that its ground plane is facing downwards. 3) Next, we introduce our bottom substrate underneath the top substrate such that their ground planes are facing each other. 4) Finally, we bond the two substrates together with a low-$\epsilon_{r}$ dielectric in between. The dielectric is highlighted with a light blue color for visualization purposes.}
\label{fig: flip-chip-scheme}
\end{figure}

\begin{figure}[tb]
\centerline{\includegraphics[width=0.5\textwidth]{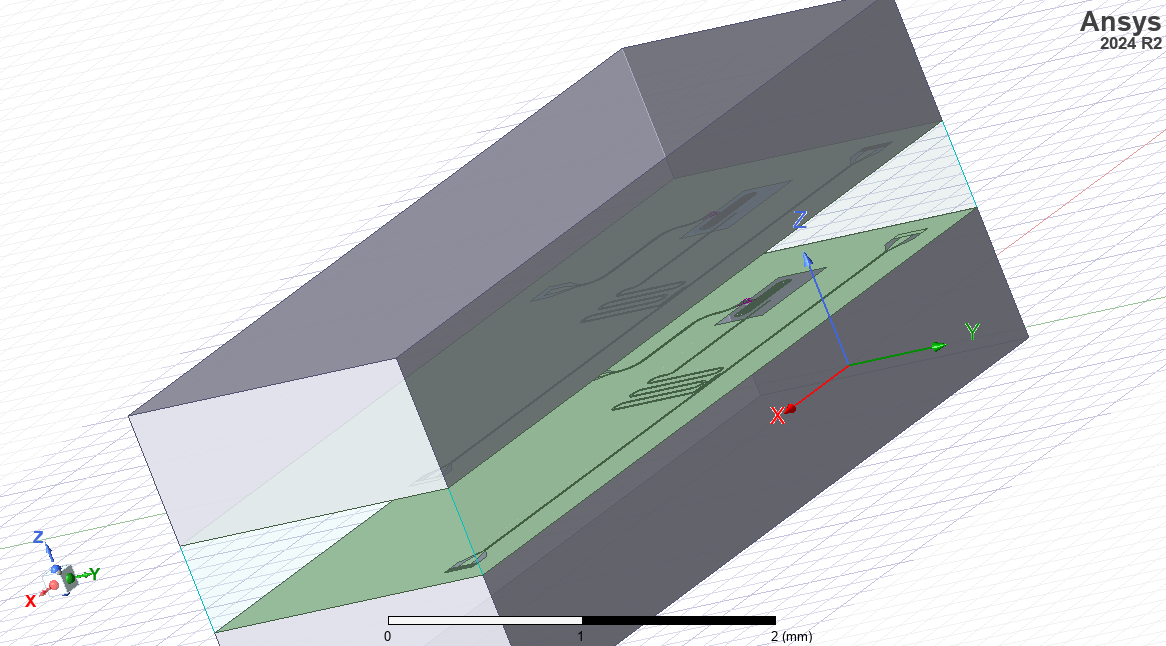}}
\caption{The 3D integrated chip used in this study with a dielectric thickness of $0.5 \text{mm}$.}
\label{sandwich-structure}
\end{figure}

In this study, we assume that two superconducting qubit circuits are fabricated on separate high-resistivity silicon substrates (Fig.~\ref{fig: flip-chip-scheme}). These substrates are then bonded using a flip-chip technique with a low-permittivity dielectric interlayer (assumed to have a relative permittivity, $\epsilon_{r}$, of 1) positioned in between them. Fig.~\ref{fig: flip-chip-scheme}  demonstrates this scheme. There are two parameters to be varied in this study, namely the dielectric interlayer thickness (or the distance between the two chips) and the quality of the dielectric (in terms of loss tangent). Unless it is specified, the dielectric is assumed to be lossless with a thickness of $0.5 \text{mm}$. Fig.~\ref{sandwich-structure} shows the 3D-integrated chip used in this study.

\begin{figure}[tb]
\centerline{
  \includegraphics[width=0.5\textwidth]{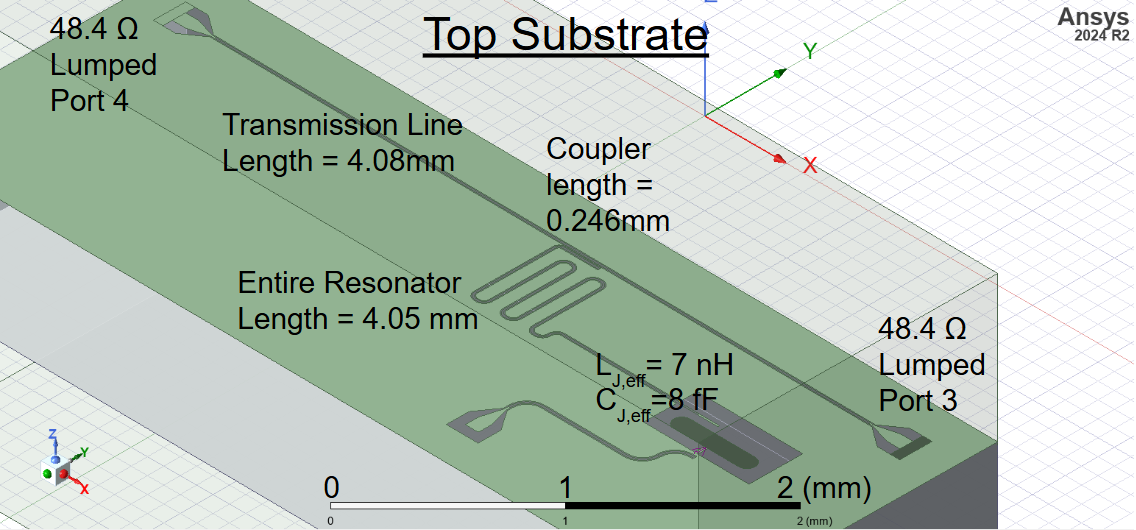}
}
\vspace{1em} 
\centerline{
  \includegraphics[width=0.5\textwidth]{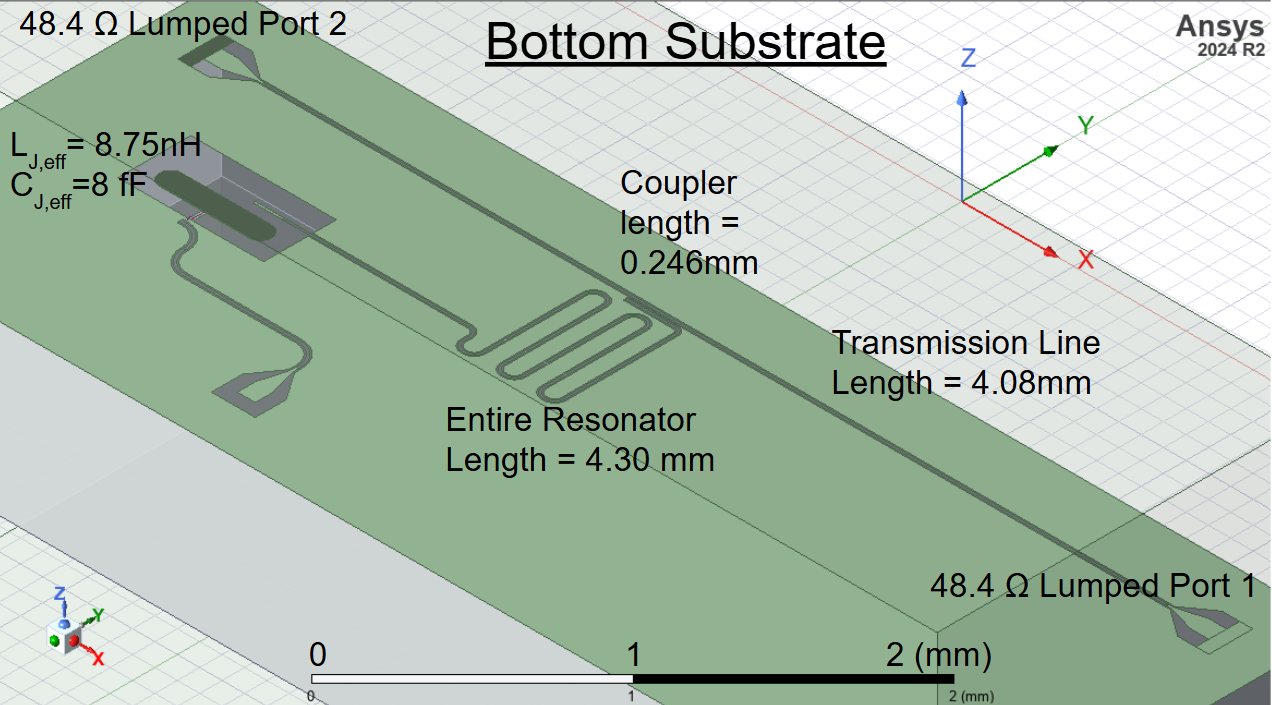}
}
\caption{Top: Top substrate and its circuit before flipping. Bottom: Bottom substrate and its circuit. Each substrate is identical in geometry with the exception of the different lengths of the quarter-wavelength resonator and the different default values of the effective Josephson inductance.}
\label{fig:device_specifications}
\end{figure}

The superconducting qubit and circuit layout and specifications on the top and bottom substrates are shown in Fig.~\ref{fig:device_specifications}.
Each substrate contains a quarter-wavelength resonator that is inductively coupled to a transmission line and capacitively coupled to a flux-tunable transmon qubit. A nearby surface bias line is included to tune the Josephson energy of each qubit. Each transmon features a SQUID loop with two Josephson junctions (JJs), enabling magnetic flux tunability. Their effective JJ inductance and intrinsic capacitance are shown in Fig.~\ref{fig:device_specifications}.

The top and bottom chips have an identical design except that the resonators have different resonant frequencies and the JJs have different effective Josephson inductance. In the final 3D integrated structure, the two circuits are almost the mirror image of the other (except that their resonators have different lengths) (Fig. \ref{sandwich-structure}). Effectively, there will be capacitive coupling between the two transmon qubits. This represents the worst case scenario to study cross-talk between the qubits in 3D integration for a given separation.


\section{Analytical Calculations and HFSS Simulations of the 3D-integrated Chip}
\label{sec: III Analytical Calculations and HFSS Simulations of the 3D-Integrated Chip}

\subsection{Transmission Line Design}

Transmission lines are important components in superconducting qubit circuits. They are used for dispersive readout and, sometimes, for qubit manipulation. Therefore, it is very important to model it correctly. Transmission lines are usually implemented using a coplanar waveguide (CPW). Its transfer characteristics (scattering matrix elements, $S_{11}$, $S_{12}$, $S_{21}$, and $S_{22}$) need to be calculated correctly. Therefore, it is important to make sure the CPW is properly designed with the desired characteristic impedance, $Z_0$.

\begin{figure}[tb]
\centerline{\includegraphics[width=0.45\textwidth]{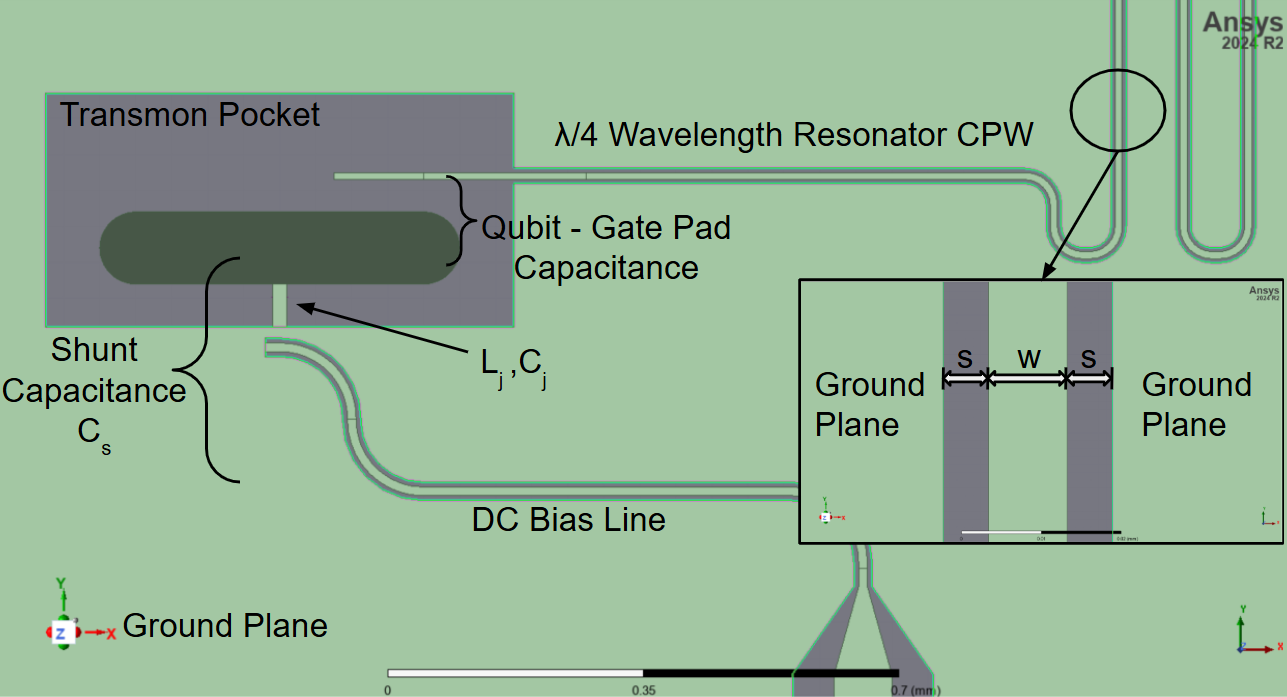}}
    \caption{Details of the transmon qubit and CPW designs. The JJ has a capacitance $C_{j}$ and inductance $L_{j}$. The transmon shunt capacitor is the capacitance between a pad and the ground plane ($C_s$).}
\label{qubit architecture}
\end{figure}

\subsubsection{Analytical Approach}

The CPW geometry (see Fig.~\ref{qubit architecture}) is defined by a trace width $w$, a trace gap $s$, and a substrate depth $H$. Since the center conductor is exposed equally to the dielectric interlayer and silicon substrate, the effective relative permittivity is approximated by the average of the two media:

\begin{align}
    \epsilon_{\text{eff}} &= \frac{\epsilon_{r,si} + \epsilon_{r,interlayer}}{2}, \\
    &= \frac{11.9 + 1}{2} = 6.45.
\end{align}

The dielectric constants are taken from the Ansys HFSS materials data table\cite{b7}.
According to \cite{b2}, the characteristic impedance of the CPW is given by:

\begin{equation}
    Z_{0} = \sqrt{\frac{\mu_{0}}{16 \epsilon_{0}\epsilon_{\text{eff}}}} \frac{K(k_{0}')}{K(k_{0})}.
\end{equation}
where $k_{0} = \frac{w}{w + 2s}$ and $k_{0}' = \sqrt{1-k_{0}^{2}}$. $\mu_{0}$ and $\epsilon_{0}$ are the free space permeability and permittivity, respectively. $K(k)$ is the complete elliptic integral of the first kind defined by:

\begin{equation}
    K(k) = \int_{0}^{\pi/2} \frac{d\theta}{\sqrt{1-k^{2} \text{sin}^{2} \theta}},
\end{equation}

Targeting $Z_0 = 50\Omega$, we chose $w = 10 \mu \text{m}$ and obtained $s = 5.806 \mu \text{m}$. By using these values and a substrate thickness of $0.75\text{mm}$ in an analytical calculator \cite{b5}, $Z_{0}$ is found to be $ = 49.568 \Omega$.


\subsubsection{Scattering Simulation Approach}

\begin{figure}[tb]
\centerline{\includegraphics[width=0.35\textwidth]{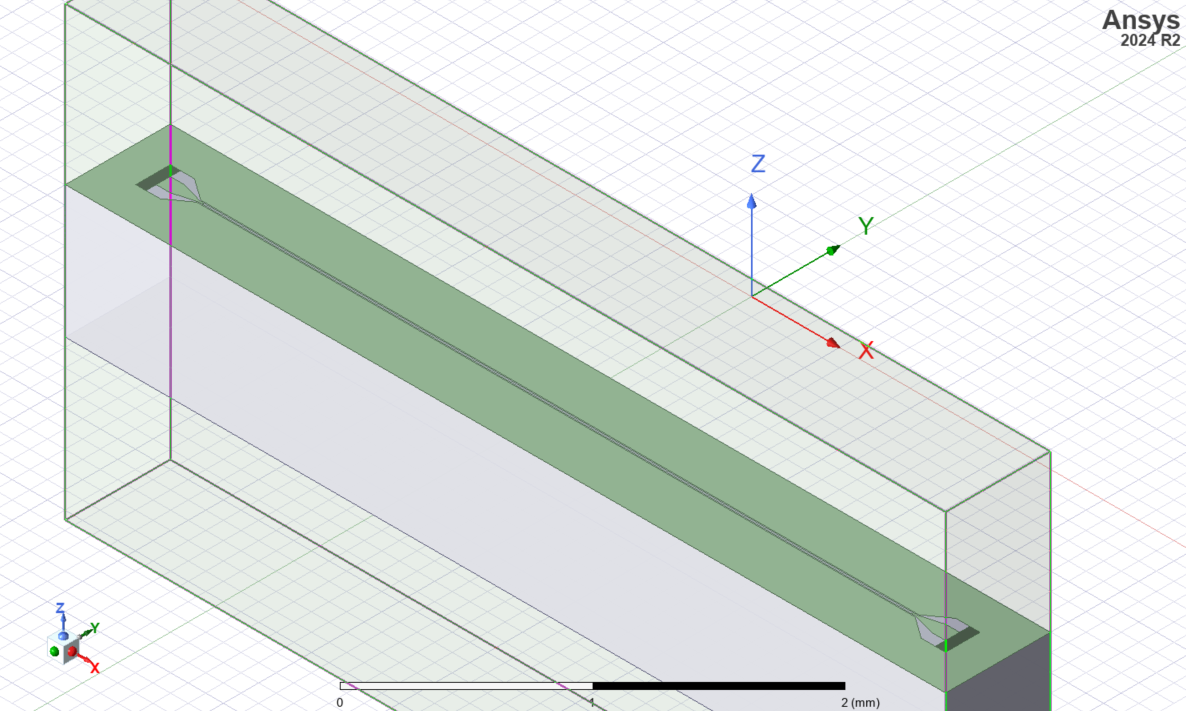}}
\caption{CPW transmission line rendering in HFSS}
\label{transmission-line-rendering}
\end{figure}

Analytical calculations involve approximations. To cross-verify our analytical results, we perform a Driven Modal simulation in Ansys HFSS, attaching lumped ports to both ends of a CPW transmission line. We then sweep over a range of port impedance values to identify the condition for impedance matching. Impedance matching occurs when the load impedance equals the transmission line impedance, resulting in minimal signal reflections within the line \cite{b1}. For each impedance value in the sweep, $Z_{\text{port}}$, we record the worse-case reflection coefficient, expressed as $\text{max[}S_{11}dB(f)](Z_{\text{port}})$. The objective of this HFSS experiment is to determine the port impedance that minimizes the worst-case reflection coefficient over our operational frequency range of $4-8 \text{ GHz}$. The structure used for finding the CPW impedance is illustrated in  Fig.~\ref{transmission-line-rendering}.

In the simulation setup, we ensure that the integration lines associated with each lumped port are oriented in the same direction along the waveguide. Additionally, we \emph{disable the impedance renormalization}, which is a common mistake made in CPW characterization.





\begin{figure}[b]
\centerline{\includegraphics[width=0.50\textwidth]{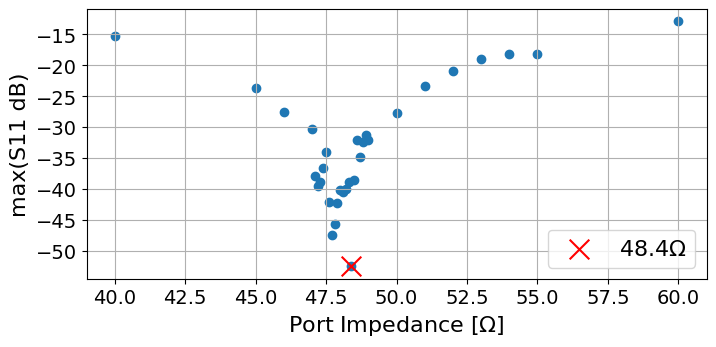}}
\caption{The worst-case reflection (max[S11 dB]) versus port impedance ($Z_{port}$). $48.4 \Omega$ is found to be the optimum value to minimize reflections.}
\label{fig:worse-case-reflections1}
\end{figure}

The result is shown in Fig.~\ref{fig:worse-case-reflections1}. It is found that the minimum worst-case reflection appears to occur at $48.4 \Omega$, which slightly differs from our analytical prediction. For the remainder of the paper, we will use $48.4 \Omega$ as the port impedance for scattering and eigenmode simulations.

\subsection{Eigenmode Simulation}

The eigenmode simulation in HFSS identifies the resonant modes of a structure, which correspond to the distinct ways in which the circuit can store electromagnetic energy. This analysis provides critical insights into the modal frequencies and field distributions of the device.
\subsubsection{Analytical Calculation}
There are four modes of interest. Two are due to the transmon qubits and two are due to the resonantors. The fundamental mode of the resonator is expected to occur at \cite{b1}

\begin{equation}
    f_{r} = \frac{v_{p}}{4l}
    \label{eq: resonator resonant frequency}
\end{equation}
where $l$ is the effective length of the resonator and $v_{p} = c/\sqrt{\epsilon_{eff}}$ is the light speed in the resonator. Since part of the resonator extends $0.25 \text{mm}$ into the transmon pocket (Fig.~\ref{qubit architecture}), which has a larger effective spacing, the effective length is expected to be between the full physical length and the length without the extension. The bottom and top resonator eigenfrequencies ($f_{r,\text{bottom}}$ and $f_{r, \text{top}}$, respectively) are estimated to be:

\begin{align}
    6.87 \text{ GHz} &<f_{r,\text{bottom}} < 7.29 \text{ GHz} \\
    7.29 \text{ GHz} & <f_{r, \text{top}} < 7.77 \text{ GHz}
\end{align}

The qubit eigenfrequencies are estimated by 
\begin{equation}
    f_{q} = \frac{\sqrt{8 E_{c} E_{J}} -E_{c}}{h}
\end{equation}
where $E_{c}$ is the \textit{charging energy} and $E_{J}$ is the \textit{Josephson Energy} given by \cite{b9}:

\begin{align}
    E_{c} &=\frac{\frac{1}{2}e^{2}}{C_{j} + C_{s}} \\
    E_{J} &= \Big{(}\frac{\Phi_{0}}{2\pi}\Big{)}^{2} / L_{j}
\end{align}
where $C_{j} = 8 \text{ fF}$, $L_{j,\text{bottom}} = 8.75 \text{ nH}$, $L_{j, \text{top}} = 7 \text{ nH}$. $C_{s}$ is designed to be $81 \text{ fF}$ (to be confirmed in Section \ref{sec : Mutual Capacitance and direct two-qubit coupling} through simulation). Solving for the qubit eigenfrequencies, we obtain

\begin{align}
    f_{q, \text{bottom}} &= 4.85 \text{ GHz} \\
    f_{q, \text{top}} &= 5.44 \text{ GHz}
\end{align}


\subsubsection{Simulation}
HFSS is used to extract the eigenmodes and the corresponding quality factor of the 3D-integrated structure. In the simulation setup, we specify a minimum frequency of $1 \text{ GHz}$ and request the solver to find $6$ eigenmodes. We configure a maximum of $12$ passes and a $1\%$ maximum delta frequency per pass. Additionally, we enable the option to Converge on Real Frequency to avoid unnecessarily strict convergence constraints. 

\begin{figure}[tb]
\centering

\begin{minipage}[b]{0.45\linewidth}
    \centering
    \includegraphics[width=\linewidth]{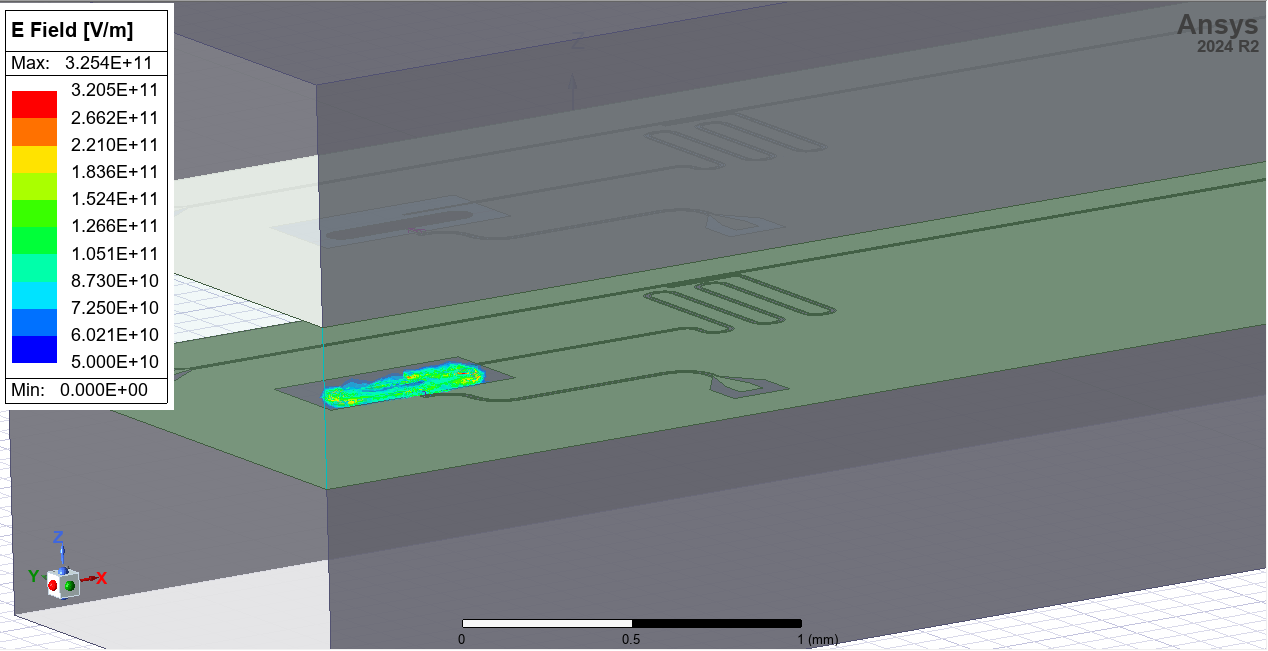}
    
    \small (a) Mode 1 : Bottom Qubit
\end{minipage}
\hfill
\begin{minipage}[b]{0.45\linewidth}
    \centering
    \includegraphics[width=\linewidth]{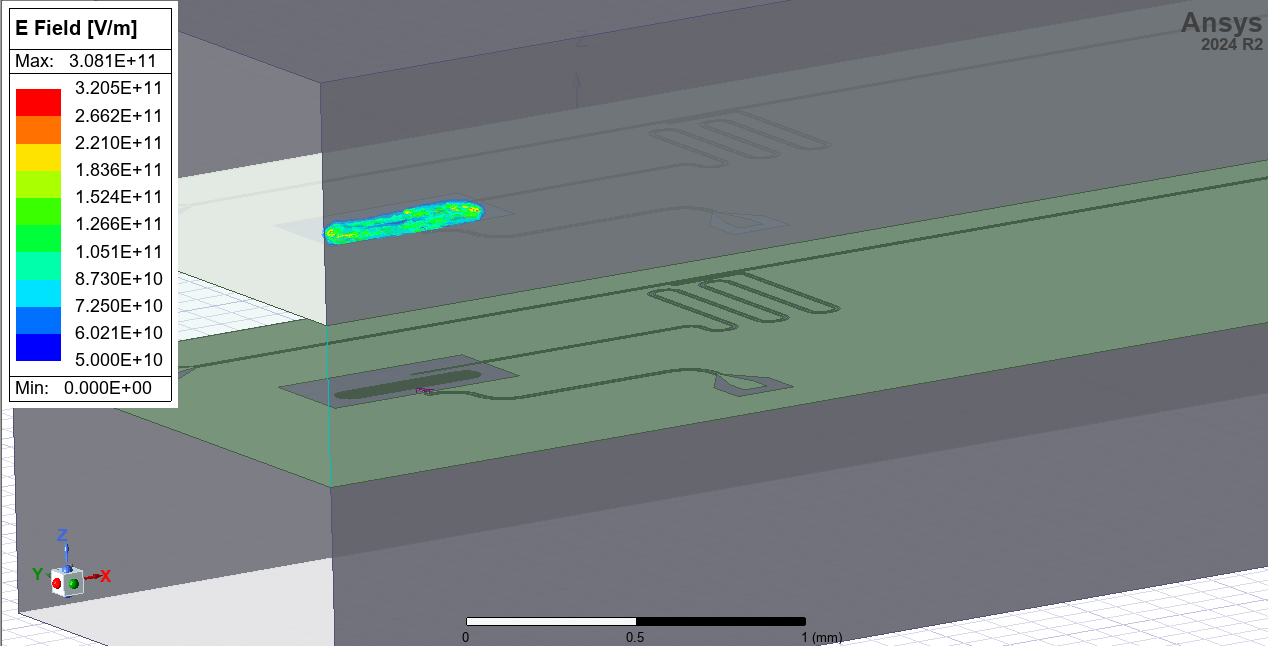}
    
    \small (b) Mode 2 : Top Qubit
\end{minipage}

\vspace{0.5em}

\begin{minipage}[b]{0.45\linewidth}
    \centering
    \includegraphics[width=\linewidth]{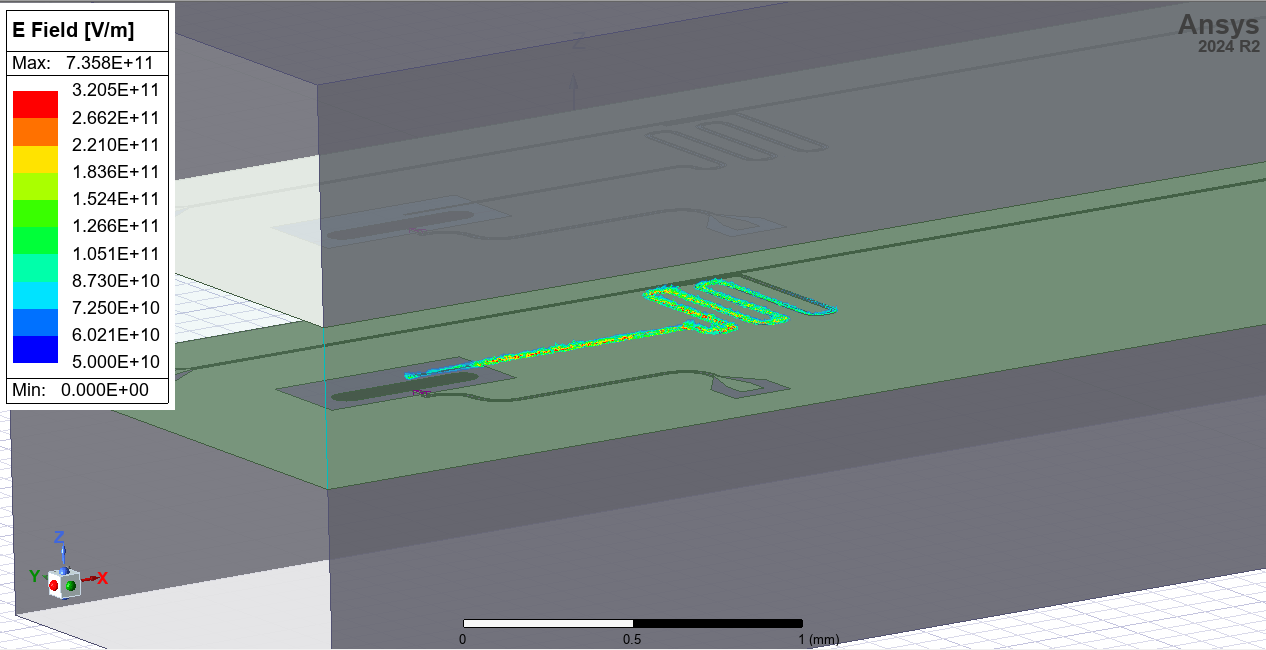}
    
    \small (c) Mode 3 : Bottom Resonator
\end{minipage}
\hfill
\begin{minipage}[b]{0.45\linewidth}
    \centering
    \includegraphics[width=\linewidth]{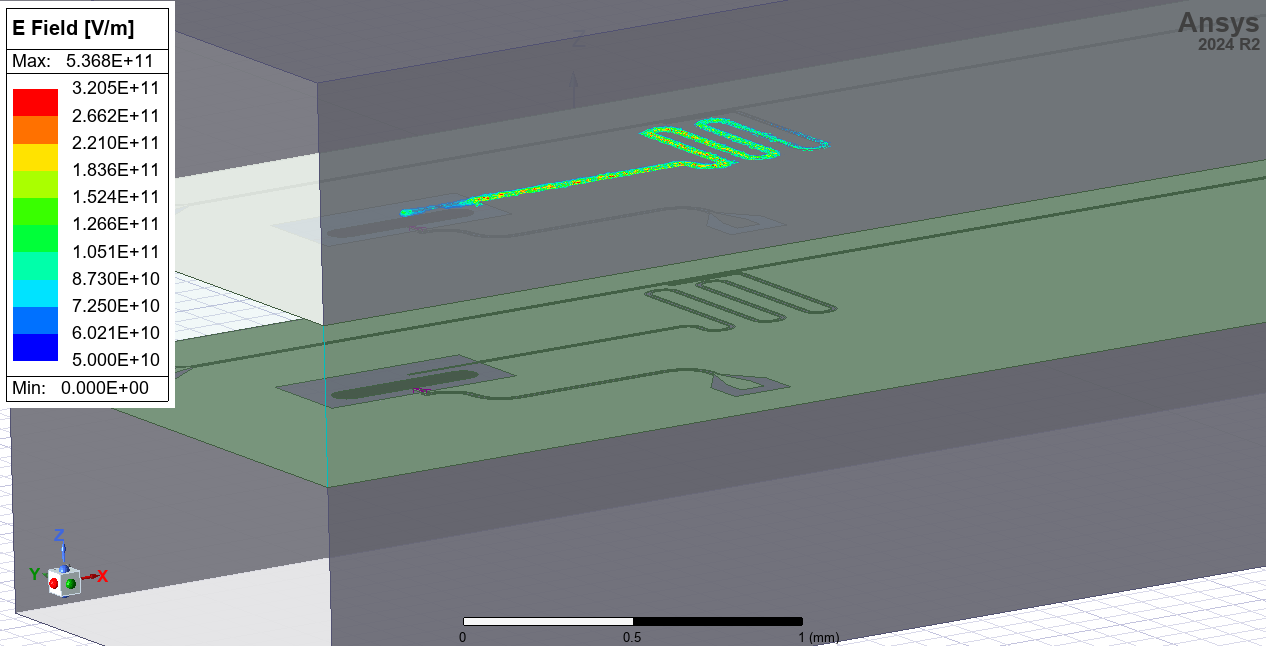}
    
    \small (d) Mode 4 : Top Resonator
\end{minipage}

\caption{Electric field distributions of the first four eigenmodes.}
\label{fig:chip_2x2}
\end{figure}

After running the simulation, we can plot the electric field distribution per eigenmode to map it to the corresponding element (Fig.~\ref{fig:chip_2x2}). Table~\ref{table:sim_vs_analytical} shows the eigenmode frequencies and quality factor of each mode. 

\begin{table}[htbp]
\caption{Comparison of Simulated and Analytical Eigenmode Data}
\begin{center}
\begin{tabular}{|c|c|c|c|}
\hline
\textbf{Eigenmode} & \multicolumn{2}{|c|}{\textbf{Simulation Data}} & \textbf{Analytical} \\
\cline{2-4}
 & \textit{Frequency (GHz)} & \textit{Q} & \textit{Frequency (GHz)} \\
\hline
Mode 1 & 5.16416 & 1.43512e+06 & 4.85 \\
\hline
Mode 2 & 5.74989 & 754259 & 5.44 \\
\hline
Mode 3 & 7.11469 & 6509.95 & 6.87 $<$ x $<$ 7.29 \\
\hline
Mode 4 & 7.50486 & 5480.00 & 7.29 $<$ x $<$ 7.77 \\
\hline
Mode 5 & 12.4485 & 607.350 & N/A \\
\hline
Mode 6 & 13.0396 + j5.74508 & 1.24011 & N/A \\
\hline
\end{tabular}
\label{table:sim_vs_analytical}
\end{center}
\end{table}

\begin{figure}[htbp]
\centering

\begin{minipage}[b]{0.70\linewidth}
    \centering
    \includegraphics[width=\linewidth]{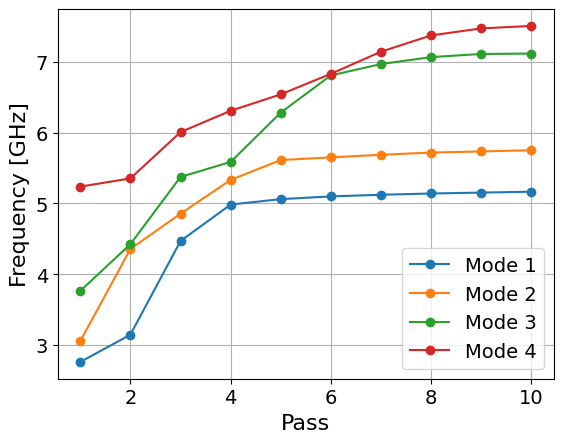}
    \small (a) Eigenfrequency vs Pass
\end{minipage}

\vspace{1em}

\begin{minipage}[b]{0.70\linewidth}
    \centering
    \includegraphics[width=\linewidth]{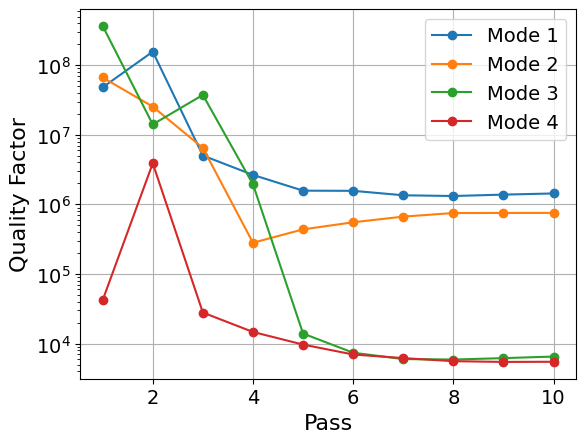}
    \small (b) $\log_{10}(Q)$ vs Pass
\end{minipage}

\vspace{1em}

\begin{minipage}[b]{0.70\linewidth}
    \centering
    \includegraphics[width=\linewidth]{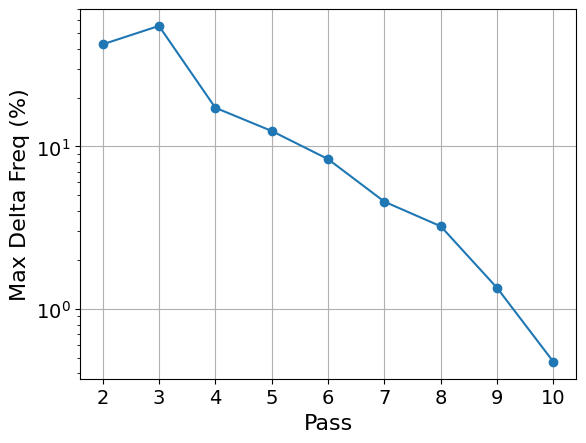}
    \small (c) Max Delta Freq \% vs Pass
\end{minipage}

\vspace{1em}

\begin{minipage}[b]{0.70\linewidth}
    \centering
    \includegraphics[width=\linewidth]{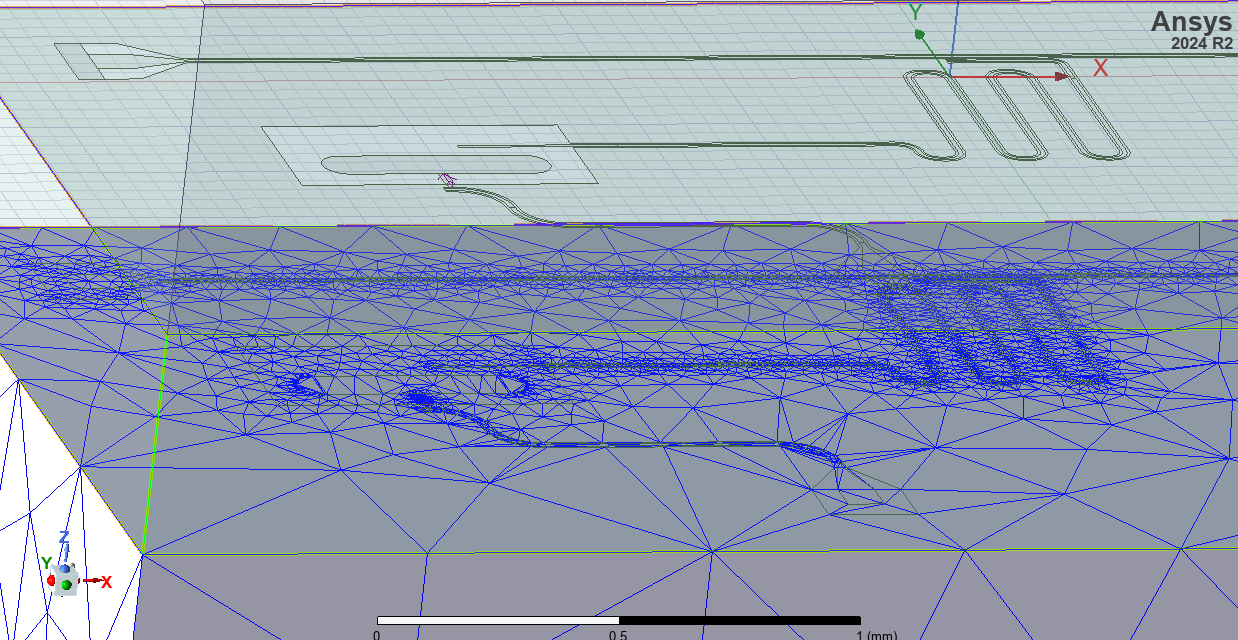}
    \small (d) Bottom Substrate Mesh
\end{minipage}

\caption{Monitoring convergence of eigenmode simulations across multiple passes: (a) Eigenfrequency trend, (b) quality factor $\log_{10}(Q)$, (c) maximum percent change in frequency, and (d) visual of the final mesh for the bottom substrate.}
\label{fig:eigenmode_convergence}
\end{figure}

To ensure the quality of the simulation, we analyze the Max Delta Freq \% vs pass, along with the eigenfrequency and quality factor versus the number of passes, as shown in Fig.~\ref{fig:eigenmode_convergence}. Qualitatively, we observe that the eigenfrequencies and quality factors of the relevant modes stabilize as the number of passes increases. The simulation converges with a maximum delta frequency percentage of $\approx 0.47\%$. Additionally, the mesh is dense in regions where electromagnetic signals are expected to propagate -- namely, the transmission line, resonator, and Josephson Junction. These simulation characteristics indicate that the simulation has converged with a sufficiently refined mesh.

\subsection{Scattering Simulation for Qubit Readout}
In scattering analysis, we expect the resonator to effectively short the transmission line at its resonant frequency, $\omega_{r}$, producing a distinct dip in the transmission response, $S_{21}$, within a certain bandwidth. However, within the framework of quantum mechanics, this dip frequency is not fixed when the resonator is coupled to a qubit. It can shift to a higher or lower frequency, depending on the qubit state \cite{b8}. This shift in frequency is known as the cross Kerr ($\chi$). When the qubit frequency, $\omega_{q}$, is less than $\omega_{r}$ (like in our case), we expect a dip at $\omega_{r} + \chi$ when the qubit is at $\ket{0}$, or a dip at $\omega_{r} - \chi$ when the qubit is at $\ket{1}$ \cite{b9}. Based on the phase change of the transmitted signal, which is dependent on $S_{21}$, the qubit state can be then inferred through this dispersive readout mechanism.

Therefore, an accurate modeling of the scattering matrix is very critical in evaluating the performance of an integrated chip. For example, will the top circuit readout $S_{21}$ be affected by the qubit state of the bottom circuit due to coupling? Moreover, scattering matrix simulation can also help verify the validity of the eigenmode simulations.

\begin{figure}[tb]
\centering

\begin{minipage}[b]{0.80\linewidth}
    \centering
    \includegraphics[width=\linewidth]{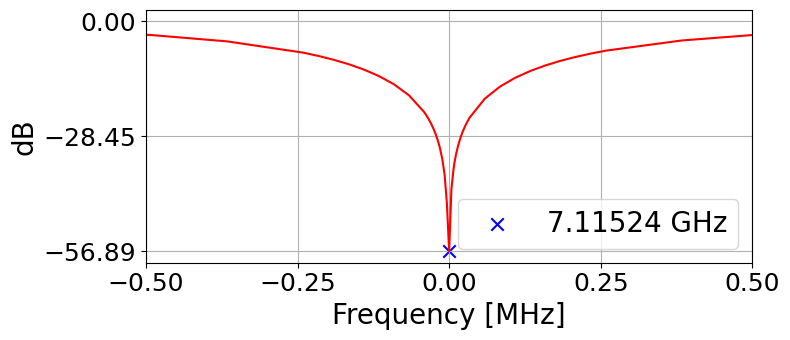}
    \end{minipage}

\vspace{1em}

\begin{minipage}[b]{0.80\linewidth}
    \centering
\includegraphics[width=\linewidth]{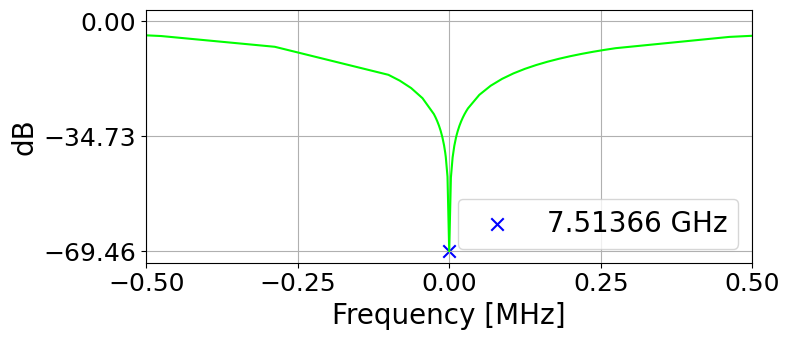}
    \end{minipage}

\caption{Top: $S_{21}$ of the bottom circuit around the resonator frequency (mode 3). Bottom: $S_{43}$ of the top circuit around the resonator frequency (mode 4).}
\label{fig:S21_charts}
\end{figure}

To ensure a successful simulation setup, we import the mesh from the eigenmode simulation into a Driven Modal simulation environment \cite{b8}. 
The imported mesh, which is already optimized for detecting resonant modes, will be further optimized for scattering parameters. After 4 passes in the Driven Modal simulation, we achieve a Max Mag. Delta S of $0.01$. Any Max Mag. Delta S less than $0.02$ is considered reasonable for most structures \cite{b15}. Then, we perform a \textit{discrete} frequency sweep over a $0.1 \text{ GHz}$ interval centered around each eigenfrequency extracted from Table~\ref{table:sim_vs_analytical}. We perform repetitive, coarse frequency sweeps, continuously narrowing our sweep window around the $S_{21}$ dips (Fig.~\ref{fig:S21_charts}).

The scattering transmission charts in Fig.~\ref{fig:S21_charts} are in agreement with the eigenmode results, predicting the resonator eigenfrequencies and their quality factors. To calculate the quality factor from the $S_{21}$ plots, we start by finding the bandwidth, $BW$, in the scattering simulations by locating the Full Width at Half Maximum (FWHM) of the scattering plots. We define the maximum as $-3 dB$. Then, we can calculate the quality factor by using:

\begin{equation}
    Q = \frac{f_r}{BW}
\end{equation}

Quality factors for the resonator modes are recorded in the table below and are consistent with the eigenmode simulation results in Table~\ref{table:sim_vs_analytical}.

\begin{table}[htbp]
\caption{Scattering Simulation Data}
\begin{center}
\begin{tabular}{|c|c|c|}
\hline
\textbf{Table} &\multicolumn{2}{|c|}{\textbf{Recorded Parameters}} \\
\cline{2-3} 
\textbf{Scattering} & \textbf{\textit{Frequency (GHz)}} & \textbf{\textit{Q}} \\
\hline
Mode 3 & 7.11524 & 6618.16 \\
\hline
Mode 4 & 7.51364 &  5782.30\\
\hline
\end{tabular}
\label{tab1}
\end{center}
\end{table}

\section{Effect of Dielectric Interlayer Thickness}
\label{sec: IV Effect of dielectric interlayer thickness}

With the robust simulation setups developed, we then study the effect of the dielectric interlayer thickness, which is also the separation distance between the top and bottom chips in the 3D-integrated structure. In the study, the dielectric thickness is changed from $0.1 \text{mm}$ to $4 \text{mm}$. 

\subsection{Parametric Analysis}

\begin{figure}[tb]
\centering

\begin{minipage}[b]{0.45\linewidth}
    \centering
    \includegraphics[width=\linewidth]{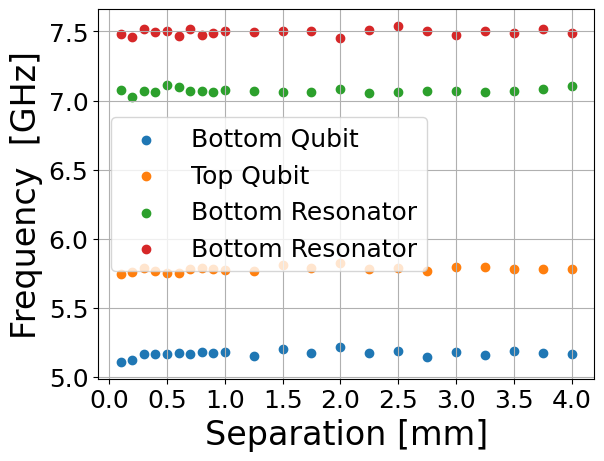}
    \\[-0.25em]
    \small (a) Eigenfrequency
\end{minipage}
\hfill
\begin{minipage}[b]{0.45\linewidth}
    \centering
    \includegraphics[width=\linewidth]{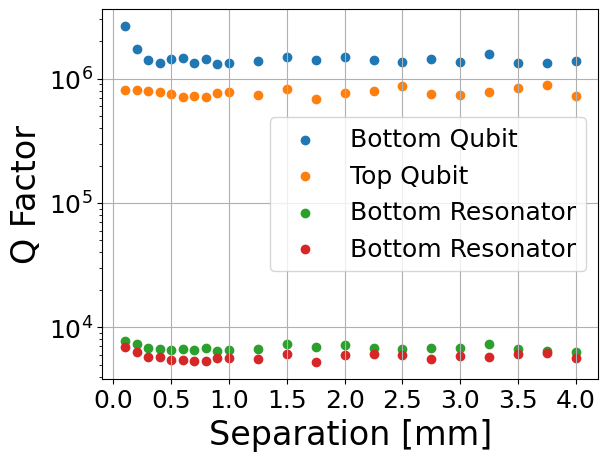}
    \\[-0.25em]
    \small (b) Quality Factor
\end{minipage}

\vspace{1em}

\begin{minipage}[b]{0.45\linewidth}
    \centering
    \includegraphics[width=\linewidth]{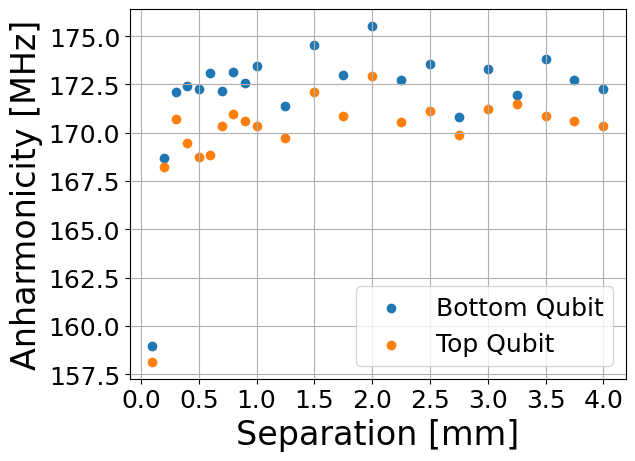}
    \\[-0.25em]
    \small (c) Anharmonicity
\end{minipage}
\hfill
\begin{minipage}[b]{0.45\linewidth}
    \centering
    \includegraphics[width=\linewidth]{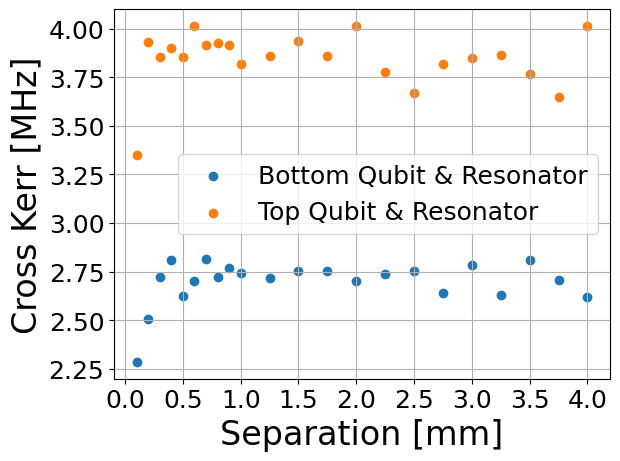}
    \\[-0.25em]
    \small (d) Cross-Kerr
\end{minipage}

\vspace{1em}

\begin{minipage}[b]{0.45\linewidth}
    \centering
    \includegraphics[width=\linewidth]{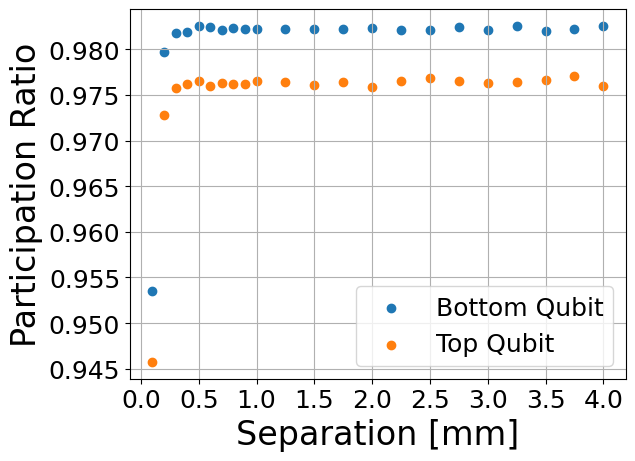}
    \\[-0.25em]
    \small (e) Energy Participation
\end{minipage}
\hfill
\begin{minipage}[b]{0.45\linewidth}
    \centering
    \includegraphics[width=\linewidth]{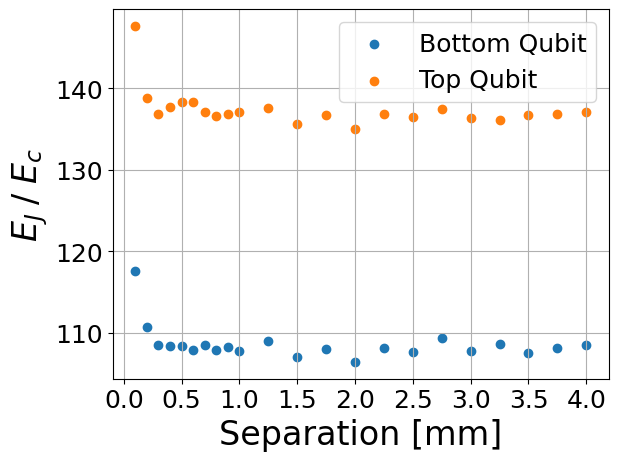} 
    \\[-0.25em]
    \small (f) $E_{J} / E_{c}$
\end{minipage}

\caption{Top and bottom circuit characteristics as a function of dielectric interlayer  thickness.}
\label{fig:chip_qualities}
\end{figure}

As shown in Fig.~\ref{fig:chip_qualities}, the HFSS eigenmode simulations and pyEPR analysis suggest that chip eigenfrequencies, Q factors, anharmonicities, and the resonator-qubit cross Kerr are constant with respect to substrate separation in the $\geq 0.5 \text{mm}$ regime. 








%



We can also estimate the upper bound of the decoherence time based on the quality factor of the qubit. 
 According to \cite{b16}, we can express the upper bound for $T_{1}$ as:

\begin{equation}
    T_{1} < \frac{Q}{2\pi f_q}
    \label{eq:T1_equation}
\end{equation}

\begin{figure}[tb]
\centerline{\includegraphics[width=0.4\textwidth]{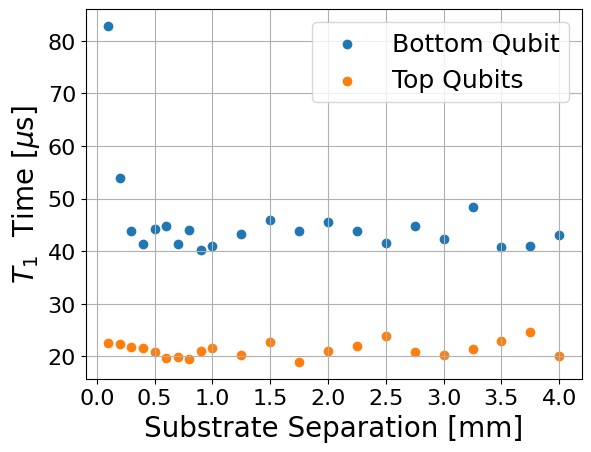}}
\caption{Upper bound limit for each qubit $T_{1}$ decoherence time. The \textit{Bottom Qubit} corresponds to \textit{Mode 1}, and the \textit{Top Qubit} corresponds to \textit{Mode 2} in our eigenmode analysis.}
\label{fig:T1 lim}
\end{figure}

Fig.~\ref{fig:T1 lim} shows the upper bound of the qubit decoherence time as a function of the dielectric thickness. It is found that the quality factor and $T_1$ upper bound increase substantially at small distance ($<0.2 \text{mm}$). The cause is not yet clear, but one possibility is due to the increase in top and bottom qubit capacitive coupling at small distance (to be demonstrated in the next Section) which reduced its resonator-mediated decoherence mechanism.

\subsection{Cross-Talk in Scattering Matrices} 

We use a scattering simulation to reveal cross-talk effects at $0.5 \text{mm}$ and $4 \text{mm}$ substrate separation by analyzing responses in the bottom chip due to signals generated in the top ports, and vice versa. 

\begin{figure}[htbp]
\centering

\begin{minipage}[b]{0.80\linewidth}
    \centering
    \includegraphics[width=\linewidth]{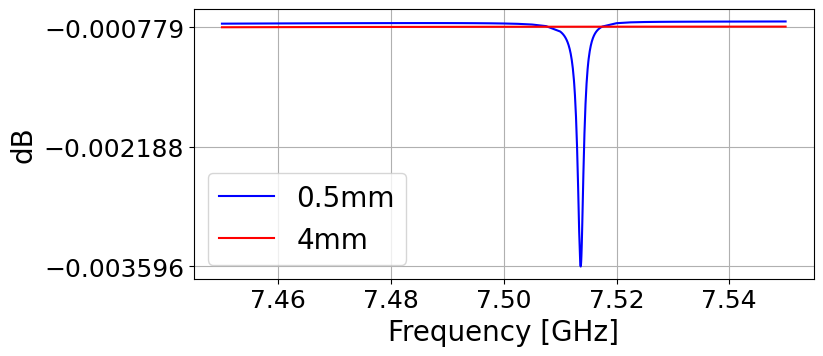}
    \small (a) $S_{21}$
\end{minipage}

\vspace{1em}

\begin{minipage}[b]{0.80\linewidth}
    \centering
\includegraphics[width=\linewidth]{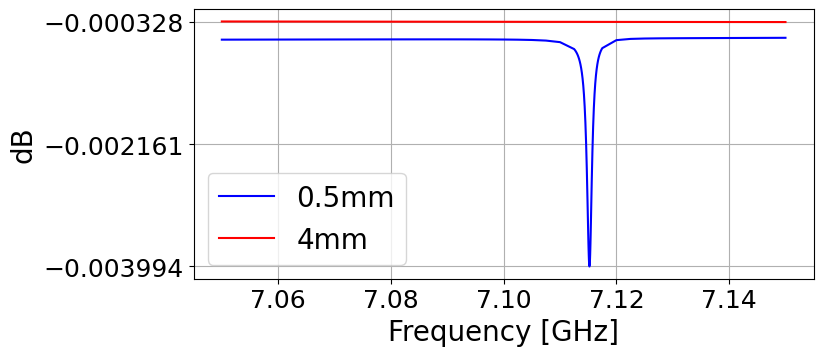}
    \small (b) $S_{43}$ 
\end{minipage}

\caption{Transmission responses for both $0.5\text{mm}$ and $4\text{mm}$ substrate separations. a) Shows a response in the top resonator due to a signal injected into the bottom substrate port. b) Shows a response in the bottom resonator due to a signal injected into the top substrate port.}
\label{fig:cross-talk}
\end{figure}

As shown in Fig.~\ref{fig:S21_charts}, when port 1 is excited, $S_{21}$ is expected to have a dip at mode 3 (the resonant frequency of the bottom resonator). Ideally, $S_{43}$ would not have this dip. Similarly, when port 3 is excited, $S_{43}$ is expected to have a dip at Mode 4 (the resonant frequency of the top resonator) and $S_{21}$ should not have a dip. Indeed, this is still true when the two chips are integrated and separated by $4 \text{ mm}$ as shown in Fig~\ref{fig:cross-talk}. 

However, when the separation is small, we observe a cross-talk in the scattering matrix simulation. At $0.5 \text{ mm}$ separation, a small dip of approximately $-0.004 \text{dB}$ is observed in $S_{43}$ (Fig~\ref{fig:cross-talk}) in response to a signal generated between ports 1 and 2. A similar dip is observed in $S_{21}$ when an excitation is applied to port 3. However, the magnitude of the dips are very small. Therefore, we may conclude that at $0.5 \text{ mm}$ separation, the top and bottom chip in the 3D-integrated structure do not interfere with each other.

\subsection{Mutual Capacitance and Direct Two-Qubit Coupling}
\label{sec : Mutual Capacitance and direct two-qubit coupling}

While the chip characteristics in Fig.~\ref{fig:chip_qualities} remain constant for detuned qubits in the $\geq5\text{mm}$ regime, the parallel plate capacitance between the top and bottom qubits varies with the substrate separation. This variation directly impacts the qubit-qubit coupling energy, $g$.

\subsubsection{Q3D Capacitance Extraction}

\begin{figure}[tb]
\centerline{\includegraphics[width=0.4\textwidth]{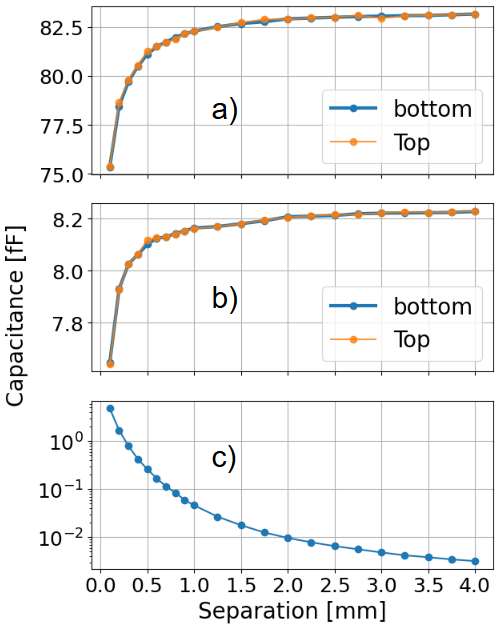}}
\caption{Substrate separation impact on qubit-related capacitance for both the top and bottom qubits. a) Shunt capacitance, $C_s$; b) the capacitance between the qubit pad and gate pad on each respective substrate; c) shows the top and bottom qubit-qubit coupling capacitance, $C_g$.}
\label{fig:q3d capacitance results}
\end{figure}

In addition to the inter-qubit capacitance, Q3D simulations  reveal that changes in the substrate separation also affect the local capacitances on each chip -- for example, the capacitance between the qubit and its neighboring gate pad, as well as the capacitance between the qubit and its shunted ground plane (see Fig.~\ref{qubit architecture} for the layout and definitions). It should be noted that $C_g$ refers to the coupling capacitance between the top and bottom transmon qubits. These relationships are summarized in Fig.~\ref{fig:q3d capacitance results}.

\subsubsection{Two-Qubit Coupling}

The coupling energy, $g$, for two directly capacitively coupled qubits in the absence of a bus resonator is given by:

\begin{equation}
    g = r \sqrt{\omega_{1}\omega_{2}}
    \label{eq: g}
\end{equation}

Here, $C_{g}$ corresponds to Fig. \ref{fig:q3d capacitance results}c, and both $C_{1}$ and $C_{2}$ correspond to the bottom and top qubits, respectively, in \ref{fig:q3d capacitance results}a. We define $r$, the \textit{capacitance ratio}, as:

\begin{equation}
    r \equiv \frac{\frac{1}{2}C_{g}}{\sqrt{C_{g} + C_{1}} \sqrt{C_{g} + C_{2}}}
\end{equation}

\begin{figure}[tb]
\centerline{\includegraphics[width=0.4\textwidth]{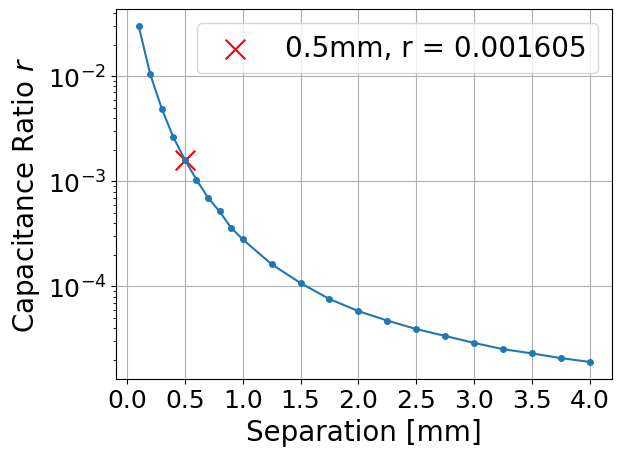}}
\caption{The substrate separation impact on the capacitance ratio, $r$, is directly proportional to the qubit-qubit coupling.}
\label{fig: Capacitance_ratio}
\end{figure}

We can track the variation of the capacitance ratio across different substrate separations in Fig.~\ref{fig: Capacitance_ratio}, which allows us to compute $g$ as a function of chip spacing. Using the capacitance ratio corresponding to a $0.5 \text{mm}$ substrate separation from Fig.~\ref{fig: Capacitance_ratio}, and the qubit frequencies $f_{q,1} = 5.16 \text{ GHz}$ and $f_{q,2} = 5.75 \text{ GHz}$ , we compute a coupling strength of approximately $g \approx 54.93 \text{ MHz}$. The two qubits on either substrate are coupled, and entanglement schemes can be applied to send quantum information between substrates.

\section{Varying the Loss Tangent of the Dielectric}
\label{sec: V Varying the loss tangent of the dielectric}

In this section, we use the same dielectric material but assign it a non-zero loss tangent to study the impact of material imperfections on qubit performance.

\begin{figure}[htbp]
\centerline{\includegraphics[width=0.4\textwidth]{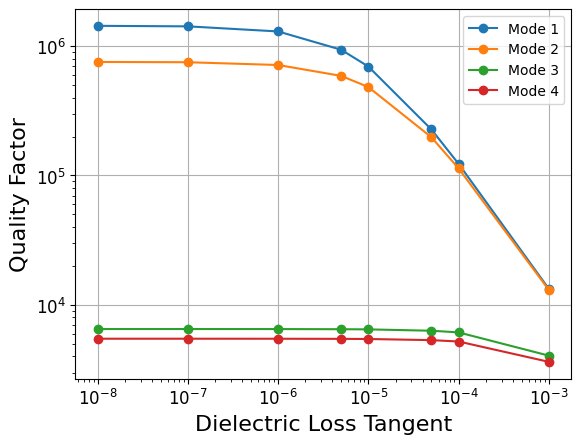}}
\caption{Q factor versus dielectric loss tangent for a $0.5 \text{mm}$ substrate separation.}
\label{fig: Qfactor_vs_DLT}
\end{figure}

For two non-coupled qubits, our eigenmode simulations in Fig.~\ref{fig: Qfactor_vs_DLT} reveal that loss tangent significantly hinders the quality factor of each resonant mode.

\begin{figure}[htbp]
\centerline{\includegraphics[width=0.4\textwidth]{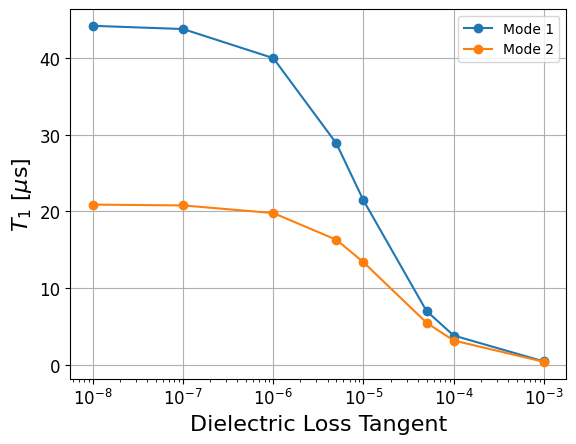}}
\caption{$T_{1}$ upper limit vs the dielectric loss tangent for $0.5 \text{mm}$ substrate separation.}
\label{fig:T1_loss_tangent}
\end{figure}

Utilizing (\ref{eq:T1_equation}), we can recast Fig.~\ref{fig: Qfactor_vs_DLT} as the upper bound on $T_{1}$ as a function of the dielectric loss tangent (Fig.~\ref{fig:T1_loss_tangent}).

\begin{figure}[htbp]
\centerline{\includegraphics[width=0.3\textwidth]{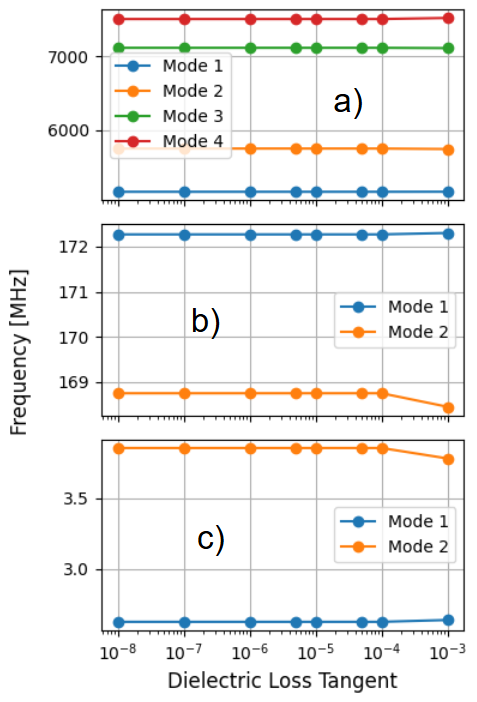}}
\caption{Quantum circuit parameters that appear to remain constant with respect to the dielectric loss tangent. The circuit metrics displayed are: a) eigenfrequencies, b) anharmonicities, and c) cross-Kerr }
\label{fig:losstangent_others}
\end{figure}

Fig.~\ref{fig:losstangent_others} shows that other circuit characteristics (extracted from eigenmode simulation and PyEPR) remain unaffected as expected.

Each qubit will also experience a decay rate induced by the dielectric \cite{b4}.

\begin{equation}
    \Gamma_{cap} = \eta_{n} \omega \sum_{i} p_{i} \text{tan} (\delta_{i})
    \label{eq:dielectric-dephasing}
\end{equation}
where $\eta_{n} \approx 1$ for a transmon, $\omega$ is the angular eigenfrequency of the qubit mode, $p_{i}$ is the electric energy participation ratio in the $i$th region of the material, and $\delta_{i}$ is the loss tangent of that material.

To calculate the stored electric energy within a specified volume, \textit{vol}, in HFSS, we begin by selecting and plotting the electric field corresponding to the eigenmode of interest. Next, we navigate to the \textit{Fields Calculator} and enter the following expression:

\smallskip

\texttt{Scl : Integrate(Volume(vol), *(Mag(<Ex,Ey,Ez>), Mag(<Ex,Ey,Ez>)))}

\smallskip

which is mathematically equivalent to the following:

\begin{equation}
    \frac{2U_{vol}}{\epsilon} = \int_{vol} |\boldsymbol{E}|^{2} d\tau
    \label{eq:energy_density}
\end{equation}

\begin{figure}[htbp]
\centerline{\includegraphics[width=0.4\textwidth]{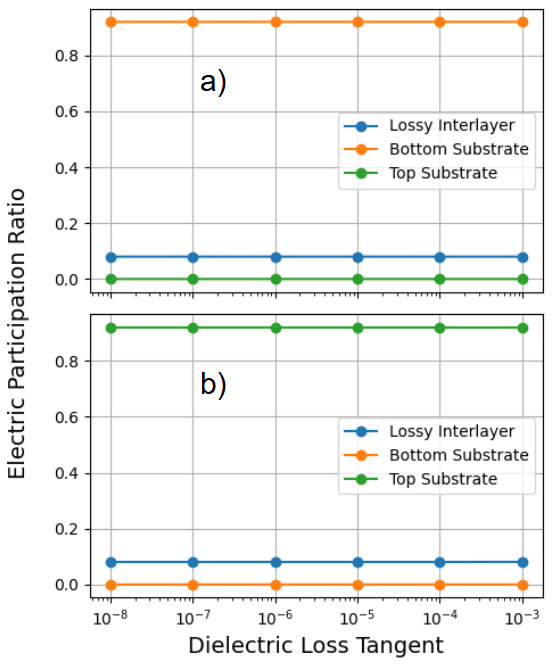}}
\caption{Electric field participation ratio of each material as a function of dielectric loss tangent for both the a) bottom and b) top qubit modes for $0.5 \text{mm}$ substrate separation.}
\label{fig:electric-participation-ratio}
\end{figure}

In Fig.~\ref{fig:electric-participation-ratio}, we plot the electric participation of the top substrate, bottom substrate, and the lossy dielectric interlayer as a function of dielectric loss tangent for both the top and bottom qubit modes.

Since the dielectric interlayer is the only material with loss tangent, it is the only material that survives in the sum in (\ref{eq:dielectric-dephasing}). Therefore, the qubit dephasing is only due to the dielectric loss tangent value and the electric energy participation ratio of the lossy material.

\begin{figure}[htbp]
\centerline{\includegraphics[width=0.4\textwidth]{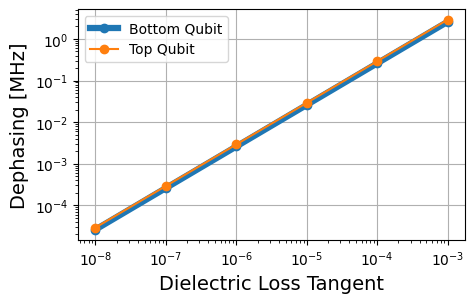}}
\caption{Qubit dephasing of the top and bottom qubit modes as a function of interlayer loss tangent for $0.5 \text{mm}$ substrate separation.}
\label{fig:Qubit-Dephasing-vs-Loss-Tangent}
\end{figure}

Fig.~\ref{fig:Qubit-Dephasing-vs-Loss-Tangent} plots the dephasing as a function of loss tangent and the trend is linear. This is because $\Gamma_{\text{cap}} \propto \tan(\delta) \approx \delta$ for small $\delta$. Moreover, Fig.~\ref{fig:electric-participation-ratio} shows that the electric participation of the interlayer remains constant with respect to dielectric loss tangent, which is consistent with the interlayer having a relative permittivity of unity. By contrast, if a lossy interlayer with relative permittivity $\varepsilon_r > 1$ is introduced (not shown), there will be polarization within the material and an increase of the electric field energy participation. This can give rise to a non-linear dependence of $\Gamma_{\text{cap}}$ on the loss tangent, in contrast to the linear behavior observed in Fig.~\ref{fig:Qubit-Dephasing-vs-Loss-Tangent}.

\section{Conclusion}

In this work, we presented a comprehensive analytical and rigorous simulation-based study of a 3D-integrated quantum chip architecture. Leveraging Ansys HFSS, Q3D, pyEPR, and analytical models from quantum theory, we've thoroughly studied the behavior, benefits, and challenges of flip-chip architecture with qubits distributed on separate substrates.

Our eigenmode and scattering simulations for detuned qubits revealed that quantities such as eigenfrequency, quality factor, anharmonicity, and the cross-Kerr shifts generally remain invariant with respect to substrate separation.

We also evaluated the impact of imperfect materials by introducing a variable loss tangent in the dielectric between the substrates. Our results show a clear degradation in $T_{1}$ relaxation times and an additional dephasing induced by the lossy dielectric. Notably, the system maintains viable performance down to a substrate separation of 0.5 mm, demonstrating the feasibility of tight 3D integration.


Overall, this work investigates the viability of 3D architectures and demonstrates that a flip-chip design is feasible and that qubit control is possible, provided dielectric losses are minimized and material quality is carefully maintained.

\section{Acknowledgement}
This material is based upon work supported by the National
Science Foundation under Grant No. 2125906.


\begin{thebibliography}{00}


\bibitem{b13} D. Rosenberg et al., “3D integrated superconducting qubits,” npj Quantum Information, vol. 3, no. 1, Oct. 2017. doi:10.1038/s41534-017-0044-0 

\bibitem{b14} D. R. Yost et al., “Solid-state qubits integrated with superconducting through-silicon vias,” npj Quantum Information, vol. 6, no. 1, Jul. 2020. doi:10.1038/s41534-020-00289-8 

\bibitem{b12} “Topologies,” D-Wave Documentation, 
https://docs.dwavequantum.com/en/latest/quantum\_research/topologies.html (accessed Apr. 8, 2025). 
\bibitem{b10} “Qiskit Metal: Quantum Device Design \& Analysis (Q-Eda) 0.1.5¶,” Qiskit Metal 0.1.5 0.1.5, https://qiskit-community.github.io/qiskit-metal/ (accessed Apr. 7, 2025). 

\bibitem{b6} Z. K. Minev et al., “Energy-participation quantization of Josephson circuits,” npj Quantum Information, vol. 7, no. 1, Aug. 2021. doi:10.1038/s41534-021-00461-8 

\bibitem{b7} Ansys, Inc., *Ansys HFSS Materials Database*, Canonsburg, PA, USA: Ansys, Inc., 2025.
\bibitem{b2} M. Göppl et al., “Coplanar waveguide resonators for circuit quantum electrodynamics,” Journal of Applied Physics, vol. 104, no. 11, Dec. 2008. doi:10.1063/1.3010859 

\bibitem{b5} “Coplanar waveguide calculator,” Microwaves101, 

https://www.microwaves101.com/calculators/864-coplanar-waveguide-calculator (accessed Apr. 4, 2025). 

\bibitem{b1} D. M. Pozar, Microwave Engineering, 4th ed. Hoboken, NJ: Wiley, 2012. 
\bibitem{b9} H. Y. Wong, Quantum Computing Architecture and Hardware for Engineers: Step by Step. Cham: Springer, 2025. 
\bibitem{b8} H. Dhillon, Y. J. Rosen, K. Beck, and H. Y. Wong, “Simulation of single-shot qubit readout of a 2-qubit superconducting system with noise analysis,” 2022 IEEE Latin American Electron Devices Conference (LAEDC), pp. 1–4, Jul. 2022. doi:10.1109/laedc54796.2022.9908196 




\bibitem{b15} Lecture 3: HFSS Fem Solution Setup, https://innovationspace.ansys.com/courses/wp-content/uploads/2021/07/HFSS\_GS\_2020R2\_EN\_LE3\_FEM\_MeshSoln.pdf (accessed Apr. 9, 2025). 
\bibitem{b16} J. C. Bardin, D. H. Slichter, and D. J. Reilly, “Microwaves in quantum computing,” IEEE Journal of Microwaves, vol. 1, no. 1, pp. 403–427, Jan. 2021. doi:10.1109/jmw.2020.3034071 

\bibitem{b4} Y. Y. Gao, M. A. Rol, S. Touzard, and C. Wang, “Practical guide for building superconducting quantum devices,” PRX Quantum, vol. 2, no. 4, Nov. 2021. doi:10.1103/prxquantum.2.040202 .
\bibitem{b3} P. Krantz et al., “A Quantum Engineer’s Guide to superconducting qubits,” Applied Physics Reviews, vol. 6, no. 2, Jun. 2019. doi:10.1063/1.5089550 





\end{thebibliography}
\end{document}